**Physics-Constrained Inverse Estimation of Irradiation-Induced Strain in He-H Ion-Implanted 4H-SiC Using Nanoindentation and Finite Element Modeling**

M. Bensalem[a]*, N. Daghbouj[a]*, J.Duchoň[b], B.S. Li[c], A.T. AlMotasem[a], S. Magalhães[d], A. Yi[e], F. Munnik[f], Xin Ou[e], W.J. Weber[g], T. Polcar[a,h]

[a]*Department of Control Engineering, Faculty of Electrical Engineering, Czech Technical University in Prague, Technická 2, 160 00 Prague 6, Czechia*

[b]*Institute of Physics of the Czech Academy of Sciences, Na Slovance 1999/2, 182 21 Prague 8, Czechia*

[c]*State Key Laboratory for Environment-friendly Energy Materials, Southwest University and Technology, Mianyang, Sichuan 621010, China*

[d]*IPFN, Instituto Superior Técnico, Universidade de Lisboa, Estrada Nacional 10, 2695-066 Bobadela LRS, Portugal*

[e]*State Key Laboratory of Materials for Integrated Circuits, Shanghai Institute of Microsystem and Information Technology, Chinese Academy of Sciences, Shanghai 200050, China*

[f]*Helmholtz-Zentrum Dresden-Rossendorf, Institute of Ion Beam Physics and Materials Research, Bautzner Landstr. 400, 01328 Dresden, Germany*

[g]*Department of Materials Science & Engineering, University of Tennessee, Knoxville, TN 37996, USA*

[h]*School of Engineering, University of Southampton, Southampton SO17 1BJ, United Kingdom*

**Abstract**

Nanoindentation is widely used to evaluate the mechanical properties of irradiated materials, however, its potential for quantifying irradiation-induced subsurface strain remains underexplored. In this work, an integrated experimental-numerical framework based on a physics-constrained inverse modeling approach is employed to estimate the magnitude of a depth-dependent irradiation-induced strain distribution in single-crystal 4H-SiC following sequential He and H ion implantation. The approach combines depth-sensing nanoindentation, finite element modeling (FEM), and a simplex-based inverse optimization routine to calibrate a physically motivated eigenstrain profile derived from ion damage simulations.

The strain field is assumed to follow a lognormal distribution consistent with independently determined damage profiles (SRIM), and is implemented in the FEM model through a depth-dependent thermal expansion formulation. By minimizing the squared error between simulated and experimental force-displacement curves, the peak tensile strain is estimated to be ~0.91%, accompanied by an effective Young's modulus of 310 GPa and a yield strength of 16.4 GPa.

Independent validation using nano-beam precession electron diffraction (N-PED) confirms good agreement between the reconstructed and experimentally measured out-of-plane strain profiles in both magnitude and spatial distribution. The results demonstrate that nanoindentation, when combined with physics-based inverse modeling, can provide a practical tool for quantifying irradiation-induced strain and residual stress in nuclear ceramics. This methodology offers a complementary approach to diffraction-based techniques for assessing subsurface damage in ion-irradiated materials relevant to advanced nuclear systems.

* Corresponding authors: e-mail: mohamed.bensalem@cvut.cz ; daghbnab@fel.cvut.cz

## I. INTRODUCTION

In advanced fusion nuclear reactors, structural components such as first walls, breeder blankets, and plasma-facing materials are continuously exposed to high-energy neutrons, ions, and transmutation products. This intense irradiation environment induces complex microstructural damage, leading to changes in mechanical properties, dimensional stability, and long-term structural integrity [1-3]. Over time, this environment leads to profound microstructural transformations [4, 5]. Under such extreme conditions, preserving the mechanical integrity, operational safety, and long-term performance of these materials represents a critical challenge. Thus, materials with exceptional properties for the extreme environment of irradiation and high temperature must be used. Among ceramic materials, silicon carbide (SiC) stands out as one of the most promising for use as a structural component in a fusion environment. Its high thermal conductivity, corrosion resistance, and radiation tolerance make it an attractive candidate for cladding and structural applications.

When neutrons or heavy ions displace atoms from their lattice sites, they generate a complex array of defects: vacancies, interstitials, dislocation loops, voids, bubbles, and stacking faults [6-14], which evolve and interact, significantly altering the mechanical integrity of the material [15, 16]. In addition, these processes are often followed by macroscale effects, such as cracks, blistering, exfoliation, hardening, and embrittlement [17-20]. A main contributor to irradiation damage is helium and hydrogen production by transmutation reactions or their introduction into the plasma-facing environment. These phenomena are closely linked to the build-up and evolution of internal strain and stress fields, which are the primary driving forces for defect growth, coalescence, and material failure. Understanding and quantifying irradiation-induced strain and stress fields in candidate materials is therefore critical for predicting their performance and lifetime under reactor-relevant conditions.

Strain in irradiated materials has traditionally been evaluated using high-resolution X-ray diffraction (XRD) and transmission electron microscopy (TEM), which provide atomic-scale insight into lattice distortions and defect structures [21-26]. While effective, these techniques are costly, time-intensive, and limited in capturing subsurface or volumetric strain distribution. Furthermore, despite the broad adoption of these methods, few studies have explored alternative pathways for strain quantification in ion-irradiated materials using non- or semi-destructive approaches.

Nanoindentation is widely used to assess irradiation-induced hardening in nuclear materials; however, its potential for quantitative reconstruction of subsurface strain fields remains largely unexplored. Because indentation response is sensitive not only to intrinsic mechanical properties but also to pre-existing residual stress, it offers a promising pathway for indirect strain quantification when combined with physics-based modeling. To address these challenges, we propose an integrated experimental-numerical framework that combines depth-sensing nanoindentation, finite element modeling (FEM), and a simplex-based inverse optimization scheme to estimate irradiation-induced strain fields in ion-implanted ceramics. While inverse FEM approaches have been widely applied to extract mechanical properties from nanoindentation data, their application to quantitative reconstruction of subsurface irradiation-induced strain profiles, validated independently by high-resolution strain mapping techniques, remains limited, particularly for brittle ceramics such as SiC. The present work extends existing inverse indentation methodologies by incorporating a physically motivated strain profile derived from ion-damage distributions and by validating the strain estimation against nano-beam precession electron diffraction (N-PED). This work establishes a physics-informed inverse methodology for estimating the magnitude of irradiation-induced strain fields, constrained by independently characterized damage profiles and validated using N-PED measurements.

## II. MATERIALS & Methods

### A. Material and Irradiation Conditions

The investigated material was a commercial single-crystal 4H-SiC wafer purchased from Goodfellow Ltd. The crystal orientation was (0001), with the c-axis normal to the polished surface. All irradiations and nanoindentation experiments were performed on the (0001) surface. The sample was first irradiated with helium (He) ions at an energy of 110 keV to a fluence of $1\times10^{16}$ ions/cm² at an elevated temperature of 800 °C. Subsequently, hydrogen (H) ions were implanted at 130 keV to the same fluence of $1\times10^{16}$ ions/cm² at room temperature. Sequential implantation with helium and hydrogen was intentionally employed to replicate key aspects of neutron-induced transmutation damage in nuclear environments. Helium implantation at elevated temperature (800 °C) promotes defect mobility, clustering, and partial dynamic annealing, mimicking in-reactor conditions. Subsequent hydrogen implantation at room temperature enables hydrogen retention and interaction with pre-existing helium-stabilized defect structures. This sequential protocol reproduces more realistic defect populations than single-ion irradiation and has been widely adopted to study synergistic He-H effects in nuclear materials [27, 28].

The average displacement damage in the top 800 nm of the sample was estimated using the Stopping and Range of Ions in Matter (SRIM 2013) code [29], employing full-cascade simulations [30] with displacement threshold energies of 20 eV for carbon and 35 eV for silicon [31, 32]. These threshold energies correspond to the minimum energy required to permanently displace a target atom from its lattice site. The displacement damage, expressed in displacements per atom (dpa), was calculated as

$$dpa = \frac{D\,\varphi}{\rho at} \tag{1}$$

where D is the total number of total displacements (Si vacancies + $C$ vacancies + replacement collision events) per unit length, φ is the total fluence, and $\rho_{at}$ the atomic number density of the material. This implantation fluence was carefully selected to achieve a peak damage level of approximately 0.18 displacements per atom (dpa), below the amorphization threshold of 0.4 dpa reported in previous studies [33, 34]. The depth profiles of He and H ion distributions and their associated dpa values were validated using Elastic Recoil Detection Analysis (ERDA). The samples were analyzed using ERDA with a 43 MeV $Cl^{7+}$ ion beam. The incident beam was directed at an angle of 75° relative to the sample normal, with a scattering angle of 30°. The analyzed area was approximately 2 × 2 $mm^2$. Recoil atoms and scattered ions were detected using a Bragg Ionization Chamber (BIC), which provides both energy measurement and atomic number (Z) identification of the detected particles. To specifically detect hydrogen and helium recoils, a separate solid-state detector was positioned at a scattering angle of 40°. This detector was placed behind a 25 μm Kapton foil, which serves to block scattered ions and heavier recoil atoms. However, the presence of the Kapton foil reduces depth resolution due to energy loss straggling. The ion beam dose, reported in arbitrary units, was monitored using a gold-coated rotating vane (rotating at 1 Hz) and a solid-state detector that recorded backscattered Cl ions from the gold surface. All events from the BIC, H/He detector, and rotating vane were recorded in list mode, allowing temporal analysis of changes occurring during measurement, particularly the loss of light elements such as hydrogen under ion bombardment. To assess elemental loss, especially of H and He, the list-mode data were used to generate plots of total recoil counts versus ion dose. These curves were then fitted and extrapolated back to zero dose to estimate the initial element content before significant ion-induced loss occurred. The data were analyzed using NDF version 9.6i [35]. List-mode files were processed to extract ERDA spectra for carbon (C), oxygen (O), silicon (Si), and the combined hydrogen and helium signal (H+He), along with the Cl backscattering spectrum on silicon. All spectra were simultaneously fitted using the NDF software. The analysis involves building a detailed sample model, from which simulated spectra are generated and iteratively fitted to the experimental data. The outcome includes best-fit spectra, the refined sample model, and depth concentration profiles of each element as a function of depth, directly derived from the experimental spectra.

As shown in Fig. 1, the peak concentrations of He and H were located at depths of approximately 570 nm and 670 nm, respectively, with maximum concentrations of ~0.8 at.% for both species. The projected ion ranges are slightly deeper than the peak radiation damage region. As ions penetrate the material, they progressively lose energy through both electronic and nuclear stopping processes. The maximum displacement damage generally occurs slightly before the ions come to rest because nuclear stopping, and therefore defect production, is highest while the ions still possess sufficient kinetic energy to efficiently displace atoms. Near the end of their trajectory, the ions rapidly lose energy and eventually stop, contributing little additional damage. Consequently, the projected ion range is typically located slightly deeper than the dpa peak. The implantation fluence of $1 \times 10^{16}$ions/cm$^2$ for He corresponds to ~0.17 dpa, whereas H implantation produces only ~0.01 dpa. Consequently, the overall damage profile is dominated by He-induced defects. The good agreement between the measured ERDA profiles and SRIM simulations confirms the reliability of the implantation and characterization procedures.

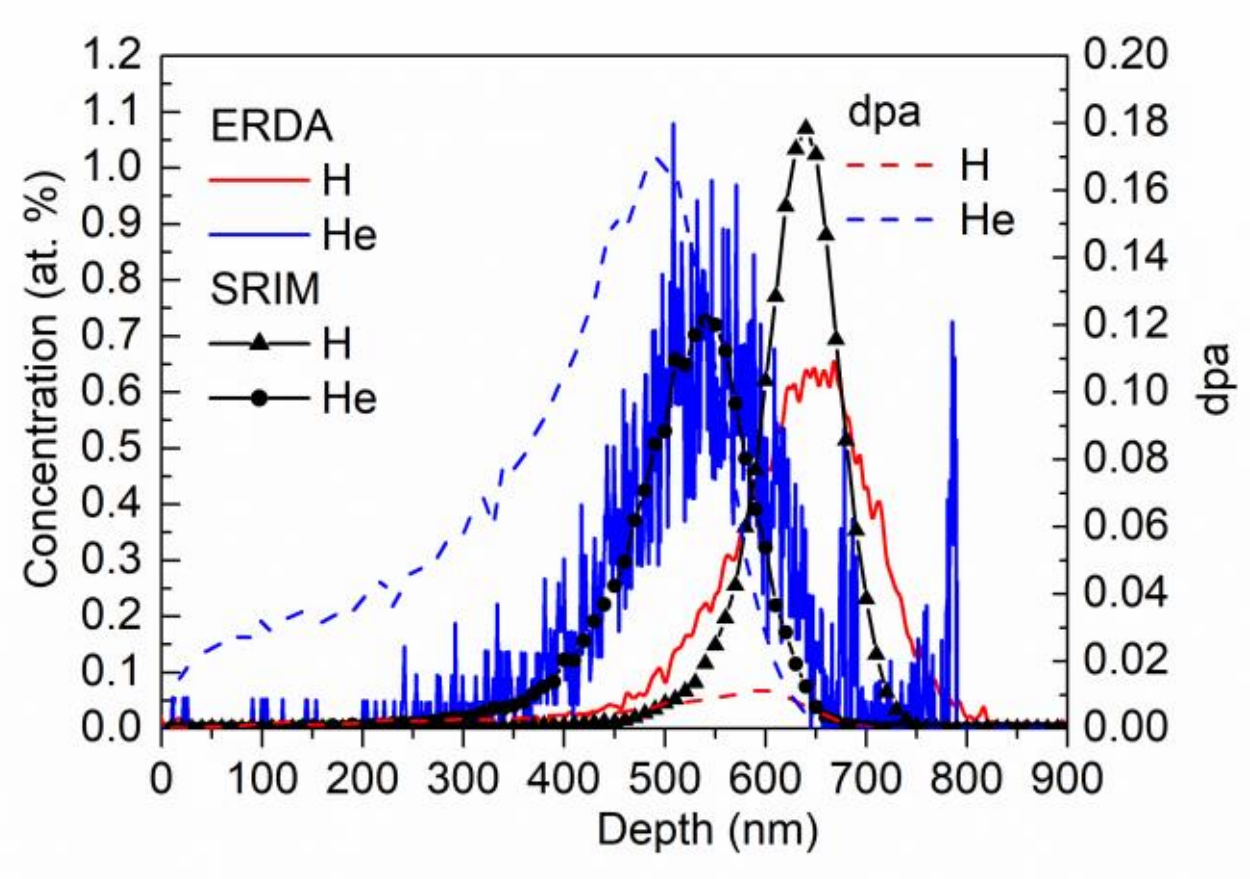

**FIG. 1.** (Left axis) Measured concentration profile obtained using SRIM simulation and ERDA in 4H-SiC after sequential irradiation with He (110 keV, $1\times10^{16}$ ions/cm² at 800 °C) and H (130 keV, $1\times10^{16}$ ions/cm² at room temperature). (Right axis) Corresponding displacement damage (dpa) simulated by SRIM.

**B. Transmission Electron Microscopy and Strain Mapping**

Transmission electron microscopy (TEM) was employed to investigate the microstructure of the 4H-SiC sample. Cross-sectional TEM samples were prepared using the FIB lift-out technique on an FEI Helios NanoLab 660 dual-beam workstation equipped with a Ga ion source operating at up to 30 kV. Initial trenches were milled at a current of 9 nA, followed by cross-sectional cleaning at 2 nA. The lamella was then extracted, transferred to a copper TEM grid, and polished using a sequential reduction of ion energy from 30 kV to 2 kV, with corresponding beam currents ranging from 20 nA to 7 pA. Subsequent imaging and strain mapping were performed using an FEI Tecnai TF20 X-Twin TEM operated at 200 kV. Both High-Angle Annular Dark-Field (HAADF) and Bright-Field (BF) imaging modes were used for structural analysis. For strain characterization, the microscope was equipped with a NanoMEGAS DigiSTAR system and controlled via the Topspin software package [36]. N-PED was employed using a 20 μm condenser aperture to produce a nearly parallel beam, yielding sharp diffraction spots suitable for quantitative analysis. In the N-PED setup, a precessed electron beam with a step size of 5 nm and a beam diameter of ~4 nm was scanned across the region of interest. Diffraction patterns were acquired at each point using a 1° precession angle and a 20 ms exposure time. This high-resolution scan allowed mapping of elastic strain fields across irradiated regions. The strain and orientation maps were generated using

Topspin's cross-correlation algorithm, which identifies diffraction spot shifts with sub-pixel accuracy. The two-dimensional elastic strain tensor was calculated by referencing each diffraction pattern to one obtained from an unirradiated, nominally strain-free region ($\varepsilon_{zz} = 0$) [37, 38], carefully selected to avoid damage artifacts. NanoMEGAS ASTAR system enabled the automated generation of orientation and phase maps with nanometer-scale resolution [39, 40]. In collaboration with AppFive, NanoMEGAS developed the "AutoSTRAIN" acquisition and analysis module, which was used here to achieve high-precision strain quantification [41-44]. The use of beam precession minimizes dynamical scattering effects and thickness-induced artifacts, enhances the visibility of diffraction spots, and improves the accuracy and spatial resolution of strain measurements [45-47]. Additionally, virtual bright-field STEM images were reconstructed by digitally placing an aperture over the central diffraction spot of each pattern, providing high-contrast visualization of the scanned regions.

**C. Nanoindentation Experiments**

The nanoindentation technique [48, 49] has been widely employed to investigate mechanical property changes at small scales, particularly those resulting from microstructural changes [50-56] or within coatings [57, 58]. The technique consists of pressing a hard indenter tip (e.g., Berkovich, Vickers, spherical, or conical), typically diamond, into the material surface under controlled loading and unloading conditions. In this study, nanoindentation was employed to evaluate irradiation-induced strain in single-crystal 4H-SiC through changes in hardness and elastic modulus after irradiation. Mechanical properties were determined from the load-displacement curves using the Oliver-Pharr method (O-P), in which the unloading stiffness is used to calculate the contact depth, contact area, hardness, and Young's modulus. To accurately probe irradiation effects, the damaged region (Fig. 1) must lie within the indentation-induced plastic zone. Therefore, indentation was first performed in quasi-static partial unload mode, consisting of 20 load-hold-partial unload cycles with a maximum load of 350 mN. This approach enables depth-dependent measurements of hardness and Young's modulus, allowing identification of the depth at which the plastic zone fully encompasses the irradiated region. Based on this analysis, quasi-static trapezoidal mode tests were subsequently performed. Nanoindentation experiments were carried out at room temperature on both virgin and irradiated samples using a diamond Berkovich indenter tip. For each sample and testing mode, 12 indents spaced 20 μm apart were performed to ensure statistical reliability and avoid overlap of neighboring plastic zones. A diamond Berkovich indenter tip with a radius of approximately 150 nm was used. The indenter elastic properties were taken as those of diamond, with $E_{ind} = 1141$ GPa, $\nu = 0.07$.

**D. Finite element modeling**

We have developed a simplified 2D axisymmetric model within ABAQUS®, which is computationally efficient and valid for reproducing the force-displacement response, as it approximates the Berkovich tip (used experimentally) as a cone with a semi-apex angle of 70.3°, thereby preserving the projected contact area. Although a Berkovich indenter produces a threefold asymmetric stress field, the 2D axisymmetric conical approximation with an equivalent projected contact area reproduces the force-displacement response with an error of approximately 3% relative to the full 3D model [57, 58]. Therefore, employing a conical indenter with a semi-apex angle of 70.3° enables equivalent indentation conditions. Since the present inverse approach relies on fitting the global force-displacement response rather than resolving local slip asymmetry, the axisymmetric approximation provides an optimal balance between computational efficiency and accuracy. Importantly, the validity of this simplification is assessed through independent experimental validation of the estimated strain using N-PED. The mechanical properties of the indenter are those of diamond ($E_{ind}$ = 1141 GPa, $\nu$ = 0.07), and the sample is assumed to behave as a perfect elasto-plastic material (no strain hardening is introduced). The boundary conditions are applied as follows: the bottom and lateral surfaces are fully fixed in the directions normal to the surfaces as well as in rotation.

The analysis consisted of two sequential steps. In the first step, irradiation-induced strain and the associated residual stress were simulated. The second step involved simulating the indentation, incorporating the residual stress field obtained from the first step as the initial state.

The formation of defects such as vacancies, interstitials, and dislocation loops is challenging to model directly using FEM, as the characteristic size of these irradiation-induced features is on the nanometer scale. Accurately resolving such small-scale phenomena requires an extremely refined mesh, which significantly increases computational cost and may introduce singularities that hinder convergence. To overcome these challenges, an equivalent strain was introduced via thermal expansion to replicate the effects of irradiation-induced strain (eigenstrain) and the resulting residual stress [59]. In this approach, the thermal expansion coefficient is defined as a function of the vertical coordinate ($y$) so that the resulting strain distribution reproduces the shape of the dpa profile. This method will be further discussed in the Results & Discussion section.

In the second step, the indentation process was simulated, and the indenter tip was introduced with a maximum penetration depth corresponding to the experimentally measured value. The displacement was controlled using an amplitude curve that describes the experimental displacement percentage as a function of time, with a strain rate of 0.05 $s^{-1}$. The mesh is composed of axisymmetric bilinear quadrilateral elements with reduced integration (CAX4R). A refined mesh was used near the contact zone, with element sizes as small as 20 nm, and gradually coarsened to 300 nm elsewhere to balance accuracy and computational time (Fig. 2). The contact between the indenter and the sample surface was modeled as surface-to-surface contact, with the indenter surface defined as the master surface and the top surface of the sample defined as the slave surface. The interaction properties were defined without specifying normal behavior, while tangential behavior was modeled using a penalty formulation with a friction coefficient of 0.2.

Preliminary studies were conducted to examine the influence of sample dimensions, the friction coefficient, and tip radius (tip bluntness). These studies showed that a friction coefficient ranging from 0 to 1 had no significant effect on the simulation results. However, the sample dimensions must be at least 7 µm (Here, a sample size of 15 µm × 15 µm was used). Furthermore, an optimization process pre-determined the effective tip radius to be 720 nm, which is in the same order of magnitude as the 809 nm found by [55]. In the present 2D axisymmetric FEM representation, the indenter effective radius was treated as an effective contact-geometry parameter used to compensate for the simplified axisymmetric approximation of the real 3D indenter geometry and for near-surface effects such as roughness and sink-in. The modeled indentation is depicted in Fig. 2. Instead of discretizing the irradiated region into artificial layers with stepwise material properties, a continuous depth-dependent eigenstrain field was introduced. This approach avoids numerical discontinuities at layer interfaces and more faithfully represents the gradual damage evolution predicted by SRIM and observed experimentally. The depth-dependent strain implicitly captures heterogeneous damage without requiring multiple material definitions.

Within this modeling framework, the use of an elasto-perfectly plastic material model represents a simplified constitutive description. This approximation is justified in the present context because the inverse analysis relies primarily on fitting the global force-displacement response rather than capturing detailed post-yield behavior. Previous nanoindentation-finite element studies on irradiated SiC and related covalent ceramics support this approach. In particular, SiC has been modeled as a perfectly elasto-plastic material, with results showing that strain hardening has a negligible influence on the simulated force-displacement response [60]. Similarly, it has been demonstrated in [61] that, for covalent materials such as SiC, the nanoindentation response can be accurately reproduced using only the Young's modulus and yield stress within a perfect elasto-plastic framework. These findings indicate that strain hardening plays a minor role in the indentation behavior of such materials. Therefore, the perfectly elasto-plastic approximation provides an appropriate and physically justified balance between model simplicity, computational efficiency, and accuracy for inverse parameter identification.

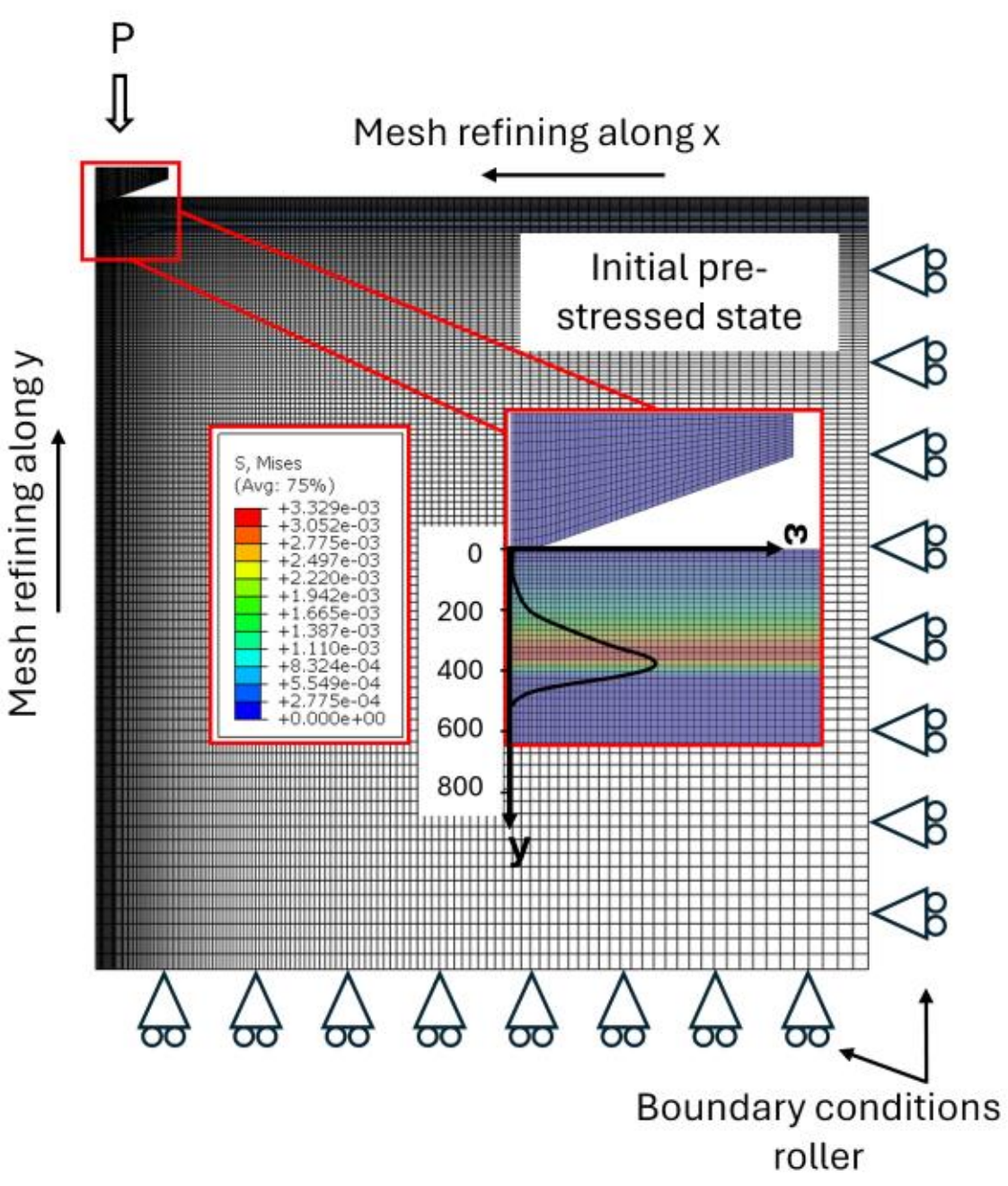

**FIG. 2.** FEM of the nanoindentation; applied boundary conditions, refined mesh with element sizes ranging from 300 nm to 20 nm, comprising a total of 21890 nodes (21270 in the sample) and 21538 linear quadrilateral elements of type CAX4R (20880 in the sample). The computation time for a single simulation is approximately 57 s for the thermal-induced strain simulation (equivalent to irradiation-induced strain) and 190 s for the indentation step, resulting in a total time of around 4 minutes per iteration. The full optimization process required approximately 4 hours.

## III. RESULTS & Discussion

### A. Irradiation-induced hardening

Figure 3 presents the averaged force-displacement curves for virgin and He+H-irradiated 4H-SiC. Each curve was generated by averaging the data from 12 independent indentations; these mean curves were then used as the primary data source for extracting mechanical properties, such as hardness, Young's modulus, and the strain amplitude.

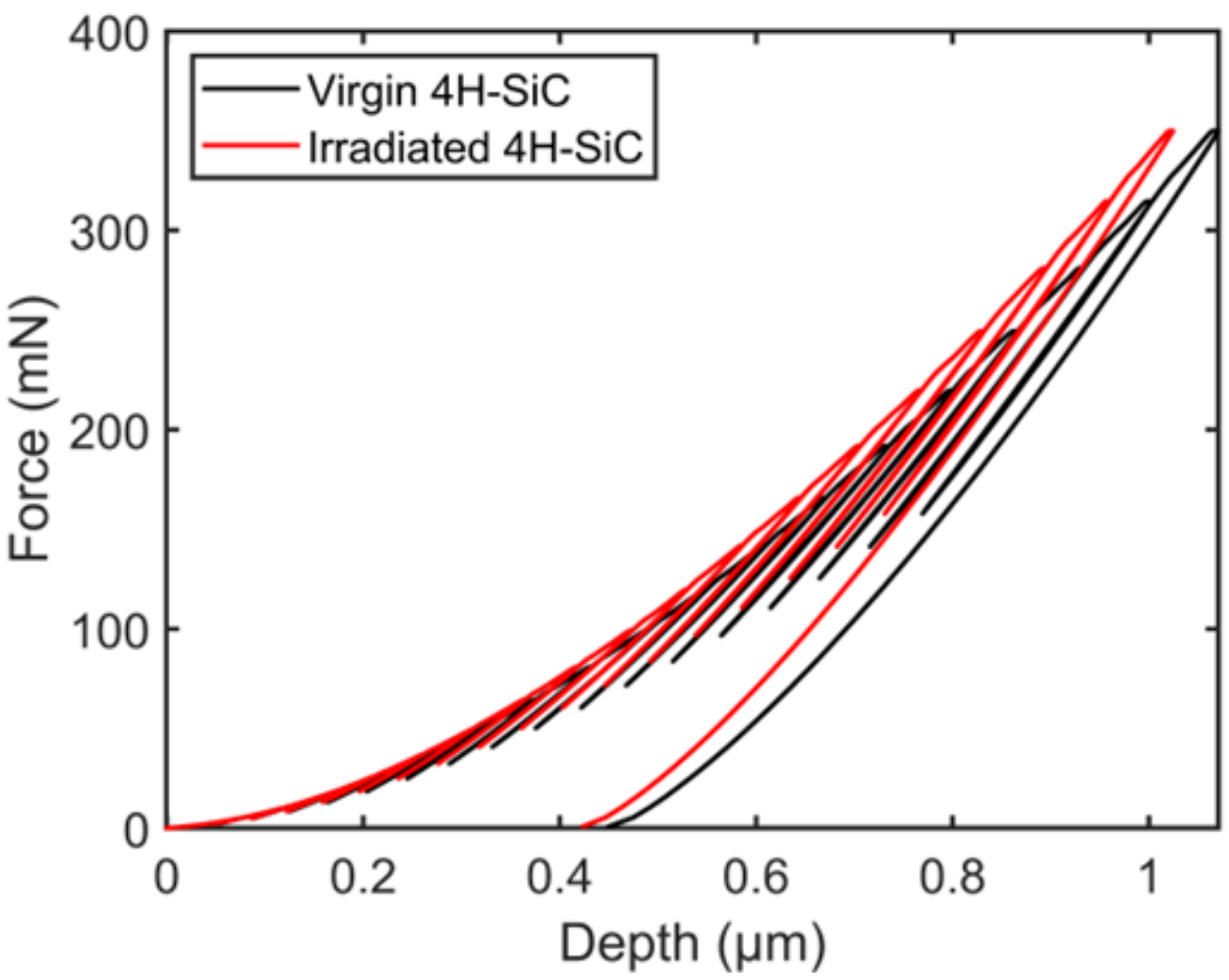

**FIG. 3.** Average force-displacement curve for virgin and He+H-irradiated 4H-SiC with $1 \times 10^{16}$ ions/cm$^2$ at a given maximum load of 350 mN and a strain rate of 0.05 s$^{-1}$.

The average force-displacement curves for both virgin and irradiated samples show a displacement shift at equivalent load values (350 mN). The irradiated sample exhibits a lower penetration depth, indicating that irradiation increased the material's resistance to indentation and thus its hardness. In order to quantify this hardness increase due to irradiation, the hardness and Young's modulus of both samples have been calculated using the O-P method for each partial unloading step and plotted versus the depth as shown in Fig. 4a. To determine a representative bulk hardness value, the data were analyzed using the Nix-Gao model, which accounts for the indentation size effect (ISE) by relating hardness to penetration depth. The Nix-Gao model describes hardness as a function of indentation depth, incorporating the characteristic length $h^*$ and the bulk hardness $H_0$ as key parameters, expressed as follows:

$$H = H_0 \sqrt{1 + \frac{h^*}{h_c}} \tag{2}$$

The fitting was performed in a region sufficiently far from the sample's surface to avoid scattering effects, while also avoiding depths where substrate influence becomes dominant, as shown in Fig. 4b. The obtained values of bulk hardness are $H_{0,irr}$ = 34.23 GPa and $H_{0,vir}$ = 32.33 GPa. Deviations at shallow depths are attributed to surface effects and residual stress, which limit the applicability of the classical Nix-Gao model in irradiated ceramics.

According to Nix-Gao, the hardness increase due to irradiation is found to be $\Delta H = H_{0,irr} - H_{0,vir} \approx 1.9$ GPa at an indentation depth corresponding to a maximum load of 35 mN. This depth corresponds to the point where the plastic zone fully encompasses the region of maximum damage, and the indentation response sensitivity to damage is highest.

The observed hardening is attributed to irradiation-induced defects, such as interstitials, vacancies, and dislocation loops, which hinder dislocation motion during nanoindentation. These defects also contribute to localized out-of-plane strain, consistent with previous findings [62], further confirming the coupled mechanical and microstructural evolution under ion implantation.

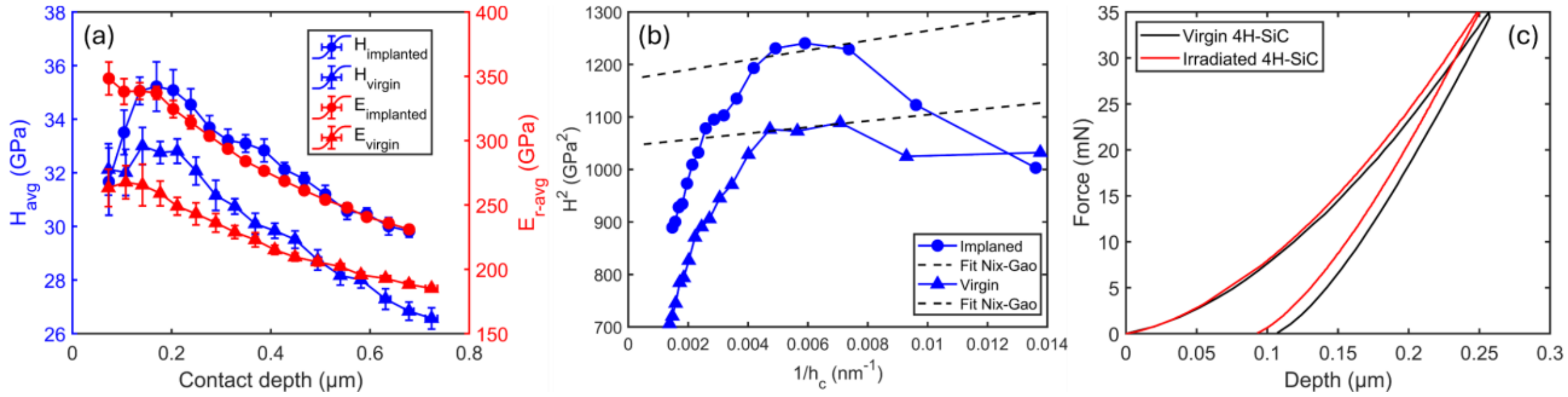

**FIG. 4.** (a) Average hardness $H_{avg}$ (left axis) and Young's modulus $E_{avg}$ (right axis) obtained using the O-P method as a function of penetration depth, (b) Nix-Gao model for virgin and He+H-irradiated 4H-SiC, (c) average force-displacement curves for virgin and irradiated samples obtained under quasi-static trapezoidal mode at a maximum load of 35 mN and a strain rate of 0.05 $s^{-1}$.

## B. Modeling and simulations

To achieve the goal of this study, namely to estimate the irradiation-induced strain along with the changes in mechanical properties resulting from irradiation, we begin by mathematically formulating the dpa profile. This approach allows the simulated profile to be expressed using constants derived from a mathematical function. The profile exhibits an asymmetric Gaussian shape, so we have tested several well-established models, including Gaussian, skewed Gaussian, lognormal, and polynomial distributions (Table S1).

The results of the fitting using the Levenberg-Marquardt algorithm in MATLAB® are shown in Fig. 5a. The choice of model is primarily based on minimizing the number of parameters describing the profile, since it is efficient to deal with as few parameters as possible to describe a function. The lognormal model provides the minimum fitting error relative to the simulated dpa profile. Table 1 shows that the lognormal model provides the best compromise between fitting accuracy and model simplicity.

**Table. 1**: Errors and Iteration numbers for the proposed fitting models.

| Model | Parameters | Error | Iteration |
|---|---|---|---|
| Gaussian | 3 | 0.0322 | 24 |
| Skewed Gaussian | 4 | 0.0088 | 37 |
| Lognormal | 3 | 0.0051 | 23 |
| Polynomial | 6 | 0.0023 | 42 |

The estimated parameters are: $c_1 = 0.1625$, $c_2 = -2.1129$, and $c_3 = 0.7359$. These parameters are respectively linked to the amplitude $A$, the position of the maximum $m$ and the standard deviation $\sigma$ as described by the following relationship:

$$\begin{cases} c_1 = A \\ c_2 = log(y_n - m) \\ c_3 = \left(\frac{1}{\sqrt{2}}\right) sinh^{-1}\left(\frac{\sigma}{e^{c_2}}\right) \end{cases} \quad (3)$$

where $y_n$ denotes the final depth point of the profile (0.618 µm), which depends on the irradiation condition. In this study, $y_n$ was fixed since the irradiation conditions were kept constant. The depth of the strain peak was fixed based on independently measured ERDA and SRIM profiles. The coefficients $A$, $m$ and $\sigma$ are given as follows: $A = 0.1625$, $m = 0.5020$ µm, and $\sigma = 0.1498$ µm.

The dpa profile represents the material damage in terms of displacement per atom, peaking at 0.18 as shown in Fig. 1. However, the primary objective is to estimate the corresponding irradiation-induced strain resulting from this damage. To evaluate the feasibility of this approach, sensitivity analyses were performed on the model's three parameters. The results indicate that while nanoindentation is sensitive to each parameter individually, the force-displacement response exhibits non-uniqueness. Specifically, similar curves can be obtained by different combinations of $c_1(A)$, $c_2(m)$, and $c_3(\sigma)$; for example, a broader distribution (higher $\sigma$) with a fixed amplitude can yield a response that is indistinguishable from a narrower distribution (smaller $\sigma$) with an increased amplitude. Furthermore, a functional dependency exists between $c_3(\sigma)$ and $c_2(m)$, complicating the simultaneous estimation of all three variables.

This highlights the intrinsic non-uniqueness of the inverse analysis when relying solely on nanoindentation data. The present approach addresses this limitation by introducing physically motivated constraints on the strain profile, derived from independent damage characterization. As a result, the inverse analysis is effectively reduced to a parameter identification problem rather than a full-field reconstruction. It is important to emphasize that the present approach does not aim to reconstruct the full strain profile independently of nanoindentation data. Instead, the profile shape is constrained using independently measured and simulated damage distributions (SRIM), and only the strain magnitude is identified through inverse analysis.

To overcome this ill-posed inverse analysis, we assume that the depth-dependent strain follows the characteristic shape of the irradiation-induced damage profile, such that both share the same peak position $m$ and standard deviation $\sigma$. Under this assumption, reconstruction of the strain field is reduced to determining a single free parameter, namely, the strain amplitude $A$. The strain is predominantly governed by the accumulation of point defects and small defect clusters, leading to local volumetric swelling. Due to limited long-range defect mobility under these conditions, the spatial distribution of strain closely follows the primary damage profile predicted by ion implantation simulations, as shown in previous studies [21, 22, 63].

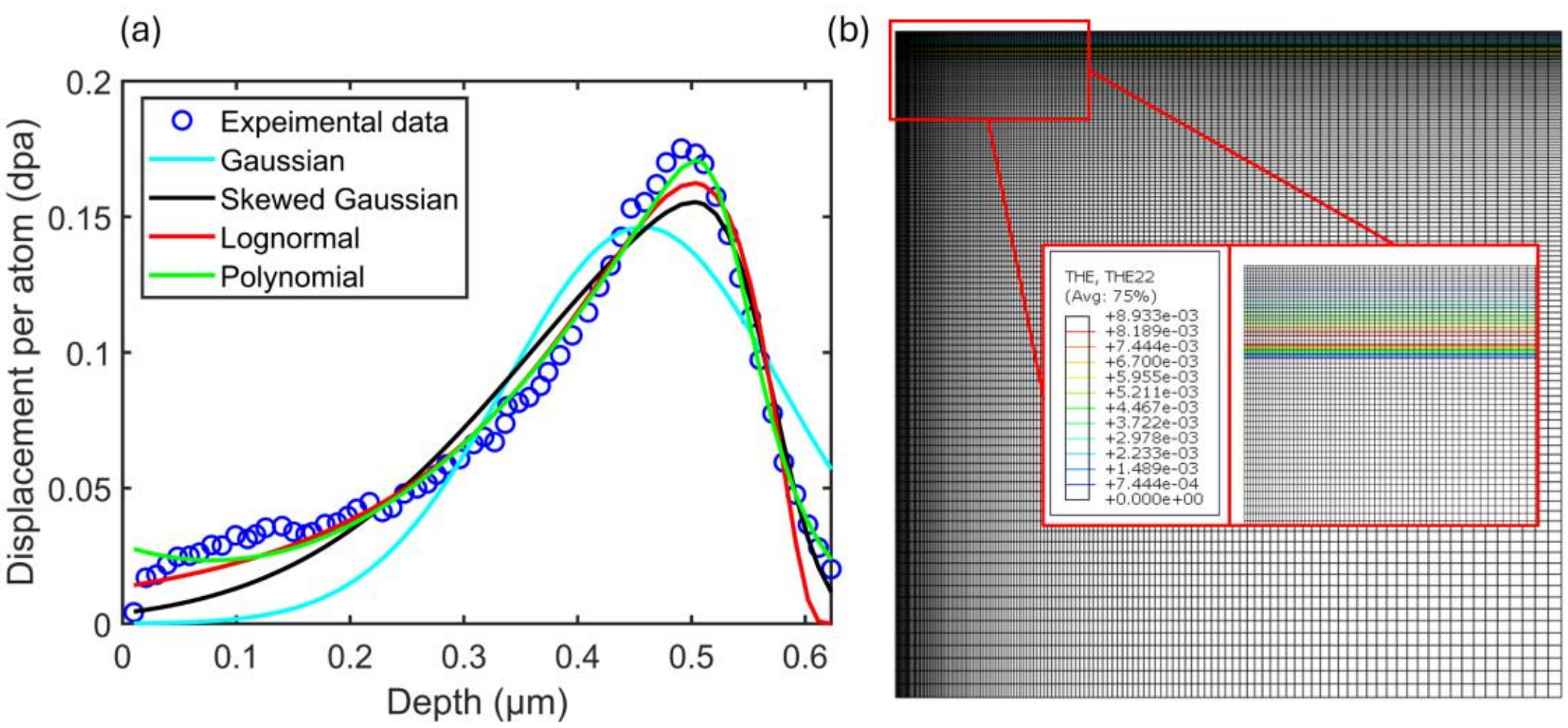

**FIG. 5.** (a) Fitting of the dpa profile obtained from SRIM using different models, (b) depth-dependent irradiation-induced eigenstrain field (THE22) generated within the finite element mesh shown in Fig. 2 using the UEXPAN subroutine prior to nanoindentation simulation..

At this stage, a sensitivity study of the amplitude with respect to the indentation response was conducted. The lognormal distribution with coefficients $c_2$ and $c_3$ obtained from the fit of the dpa profile along with various strain magnitudes was introduced, ranging from 0.1% (0.001) to 4% (0.04). The resulting strain profiles are illustrated in Fig. 6a. The strain was fed into the model using the Fortran subroutine UEXPAN, which receives the nodal coordinates from an input field containing the $y$-values and utilizes them to compute the strain increment over a time increment as a function of depth. This increment is then returned to ABAQUS® for simulation. The mathematical expression for the strain profile is defined as follows:

$$\Delta\varepsilon_{irr} \Leftrightarrow \Delta\varepsilon_{th} = \kappa \times \Delta T \tag{4}$$

$\Delta\varepsilon_{irr}$: Irradiation-induced strain increment (-).

$\Delta\varepsilon_{th}$: Equivalent thermal strain increment (-).

$\Delta T$: Temperature increment ($K$).

$\kappa$: Thermal expansion coefficient as a function of $y$.

Since the strain has the shape of a lognormal distribution along the irradiation direction ($y$), the thermal expansion should have this profile as the temperature increment is constant.

$$\Delta\varepsilon_{th} = f(y) \Longrightarrow \kappa = \kappa_0 \times c_1 \times e^{-\frac{(\log(y_n - y) - c_2)^2}{2{c_3}^2}} \tag{5}$$

$\kappa_0$ is taken as $4.5 \times 10^{-6} K^{-1}$.

In this case, $\Delta T$ is chosen as the inverse of $\kappa_0$ ($2.2222 \times 10^5$) to obtain a strain distribution controlled by the lognormal law as follows:

$$\Delta\varepsilon_{th} = \underbrace{\kappa_0 \times \Delta T}_{1} \times c_1 \times e^{-\frac{(\log(y_n - y) - c_2)^2}{2{c_3}^2}} \tag{6}$$

To improve numerical convergence, $\Delta T$ is taken to be $222.22\ K$; thus the strain is given by:

$$\Delta\varepsilon_{th} = \underbrace{\kappa_0 \times \Delta T \times coef}_{1} \times c_1 \times e^{-\frac{(\log(y_n - y) - c_2)^2}{2{c_3}^2}} \tag{7}$$

With $coef = 10^3$, where, $c_1$ is provided to the UEXPAN subroutine and $\kappa_0 \times \Delta T \times coef \times e^{-\frac{(\log(y_n - y) - c_2)^2}{2{c_3}^2}}$ is coded within it. This expression allows the irradiation-induced strain to be modeled as an equivalent thermally induced-strain, defined by a thermal expansion coefficient $\kappa$ that is a function of depth $y$.

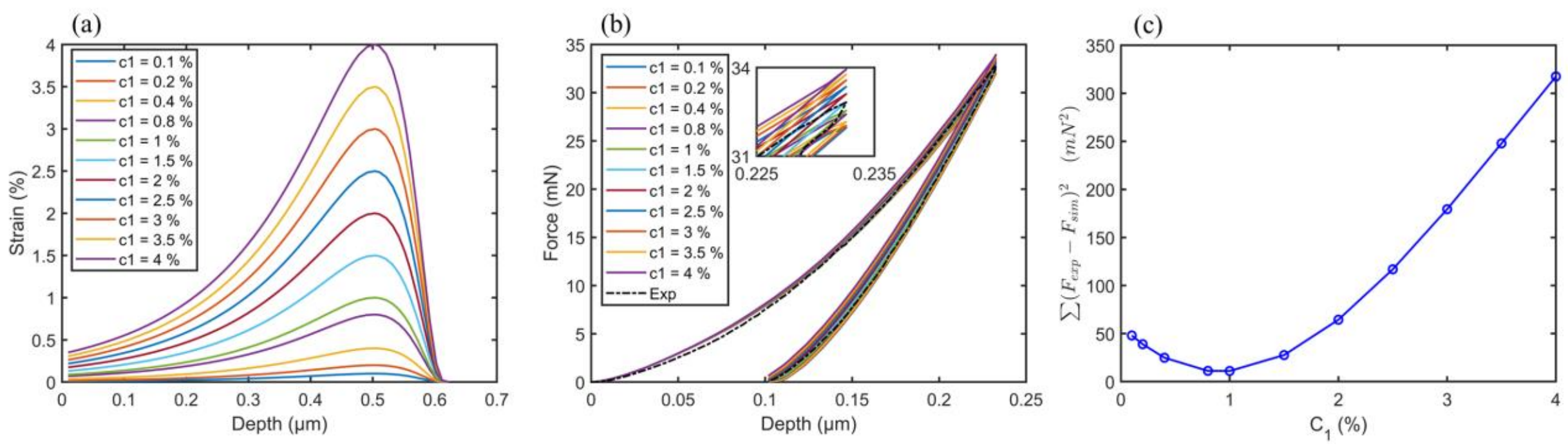

**FIG. 6.** (a) Strain profiles with different amplitudes $c_1$ ranging from 0.1% to 4% and (b) corresponding force-displacement curves obtained from FEM simulations, (c) squared error between experimental and simulated indentation force values as a function of the strain magnitude in % in the context of a sensitivity study according to the proposed model.

Once the thermal strain is generated, the resulting stress field is integrated into the mechanical model. The nanoindentation test is then simulated, and the corresponding force-displacement responses are extracted, as presented in Fig. 6b. The results show that as the amplitude $c_1$ increases, the predicted indentation response also increases. This behavior is consistent with expectations, as radiation damage introduces a tensile strain along the indentation axis, which in turn induces in-plane compressive stress (radial direction $x$) [64]. Such compressive stress enhances the material's resistance to indentation, resulting in a higher force for the same indentation depth (inset in Fig. 6b), confirming the observed increase in Nix-Gao hardness in the irradiated sample compared to the virgin sample.

A norm-based error metric is employed to compare the simulated and experimental force values (Fig. 6 c) rather than relying solely on the nanoindentation peak load [65]. The minimum error corresponds to the optimal range of $c_1$ that provides the best agreement between the simulated and experimental force-displacement data. This result underscores the importance of accurately capturing the magnitude of irradiation-induced strain when predicting the mechanical response of irradiated materials. The sensitivity analysis also indicates that the identified strain magnitude is reasonably well constrained, as variations in strain amplitude produce measurable changes in the force-displacement response, supporting the robustness of the inverse identification. A variation in the maximum strain from 0.1% to 4% results in a change of maximum force of approximately 0.002 N, corresponding to a sensitivity of 19.5 $N^{-1}$.

Optimization algorithms are essential tools in various engineering and scientific applications, including calibrating finite element models and estimating material properties. These algorithms aim to minimize or maximize a given objective function by iteratively adjusting the variables. Among the most widely used optimization algorithms are gradient-based methods, evolutionary algorithms, and direct search methods like the Nelder-Mead simplex algorithm [66].

First-order methods, also known as gradient-based methods like steepest descent and conjugate gradient, rely on calculating the gradients (or derivatives) of the objective function with respect to the parameters. These algorithms are typically fast and efficient when the objective function is smooth and differentiable. However, the main drawback is their tendency to get trapped in local minima, making them less effective when dealing with noisy, discontinuous, or non-differentiable functions.

In contrast, second-order methods, such as Newton's method and quasi-Newton methods, provide faster convergence by incorporating curvature information from the second derivative (Hessian matrix). Newton's method uses both the Jacobian and the Hessian matrices to determine the direction and size of each step, resulting in rapid convergence, especially near the solution.

However, calculating the Hessian can be computationally expensive for high-dimensional problems, and the method can become unstable if the Hessian is poorly conditioned.

In this study, a different approach, called a direct search method (also referred to as zero-order, black-box, pattern search, or derivative-free), is employed to estimate the mechanical properties of irradiated samples. Specifically, the Nelder-Mead simplex method is used [66, 67]. Unlike gradient-based methods, the Nelder-Mead method does not require any gradient or Hessian information to guide the optimization process. Instead, it uses a simplex, a geometric shape formed by evaluating the objective function at its vertices, to search for the minimum [67]. The method incrementally adjusts the simplex's shape and step size as the search progresses. Its main advantage lies in its simplicity and versatility, making it suitable for a wide range of problems, particularly those with non-differentiable objective functions or those derived from simulations where gradients are not easily accessible.

The Nelder-Mead method follows a set of rules that guide how the simplex is updated, based on the evaluations of the objective function at its vertices [67]. The process is illustrated in a flowchart in Fig. S1 and Fig. S2.

The optimization workflow shown in Fig. 7 is employed not only to estimate the irradiation-induced strain and resulting residual stress, but also to simultaneously determine the mechanical properties of the irradiated material. This is achieved through fitting FE simulation results to the experimental nanoindentation data, rather than relying on the O-P method. This is essential, as irradiation induces hardening effects. The entire process from reading the experimental data, updating the parameters in the UEXPAN subroutine using the simplex algorithm, and running ABAQUS® and monitoring the generation of output files (e.g., the odb file), is fully automated and managed through a Python script. Initially, the script reads the processed experimental data and inserts an initial guess for the mechanical properties (Young's modulus $E$ and yield strength $\sigma_{y0}$) into ABAQUS®, while assigning a trial strain magnitude within the Fortran subroutine file UEXPAN. This subroutine uses the specified strain magnitude to compute the strain increment at each iteration, which is subsequently passed to ABAQUS®. Upon completion of the first simulation (via continuous monitoring of the odb file), the script initiates a second simulation of nanoindentation, which imports the resulting stress field as predefined fields ($S_{11}$, $S_{22}$, $S_{12}$). Once the second simulation ends, the script extracts the simulated force-displacement data. To ensure consistent comparison, the simulated force is interpolated onto the experimental displacement data, and the sum of squared errors between the experimental and projected force values is evaluated against a predefined convergence criterion.

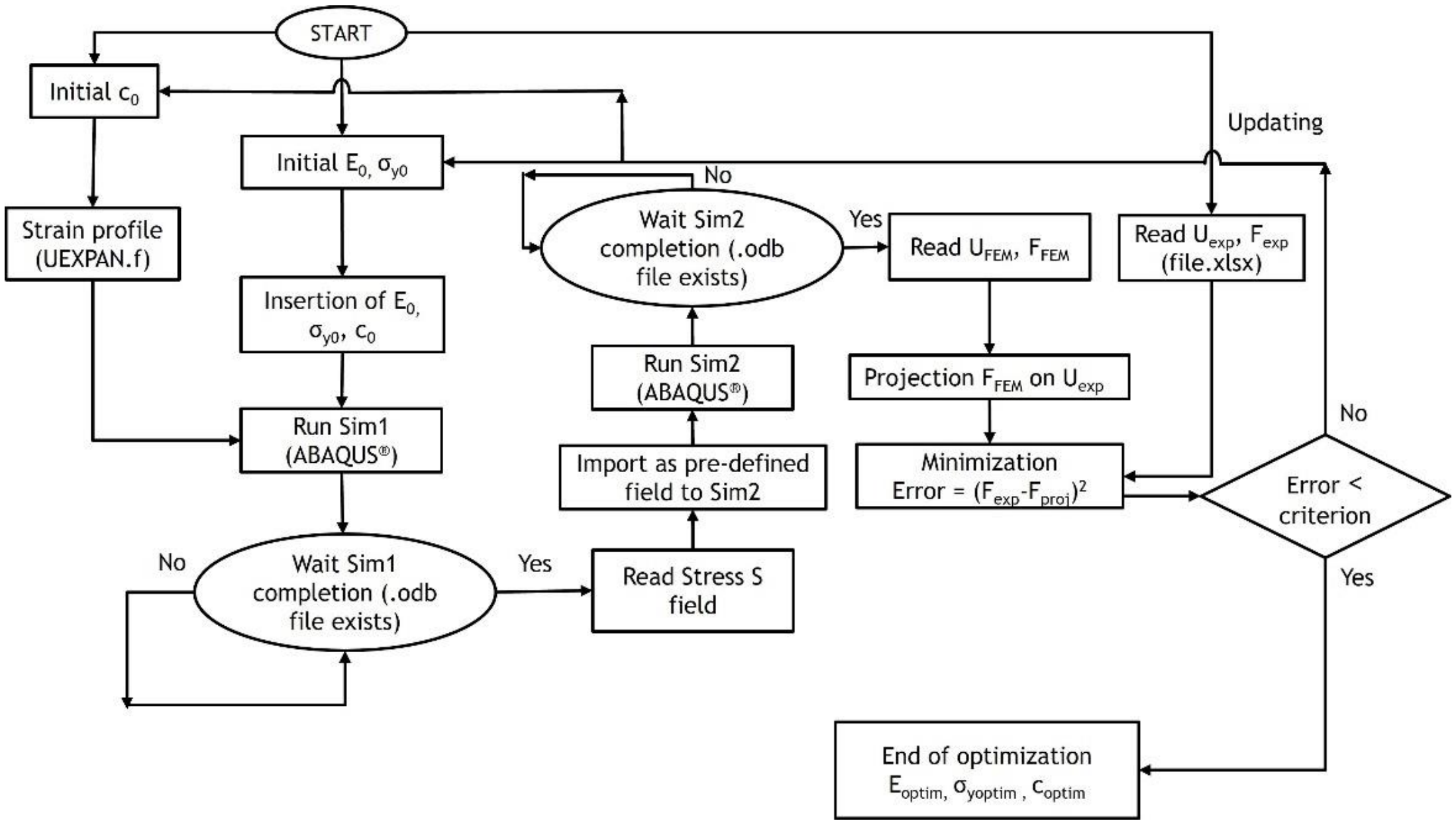

**FIG. 7.** Optimization workflow developed in Python to estimate the strain and mechanical properties of He+H-irradiated 4H-SiC through comparison between simulated and experimental nanoindentation data.

If the convergence criterion is not met, the parameters $c_1$, $E$, and $\sigma_{y0}$ are updated using the optimization algorithm, and the loop is repeated. This iterative process continues until the error falls below the defined threshold. Upon convergence, the optimization yields the final calibrated values: $c_{optim}$, $E_{optim}$, and $\sigma_{optim}$. The results of the estimation of strain magnitude and mechanical properties, along with errors, are presented in Fig. 8.

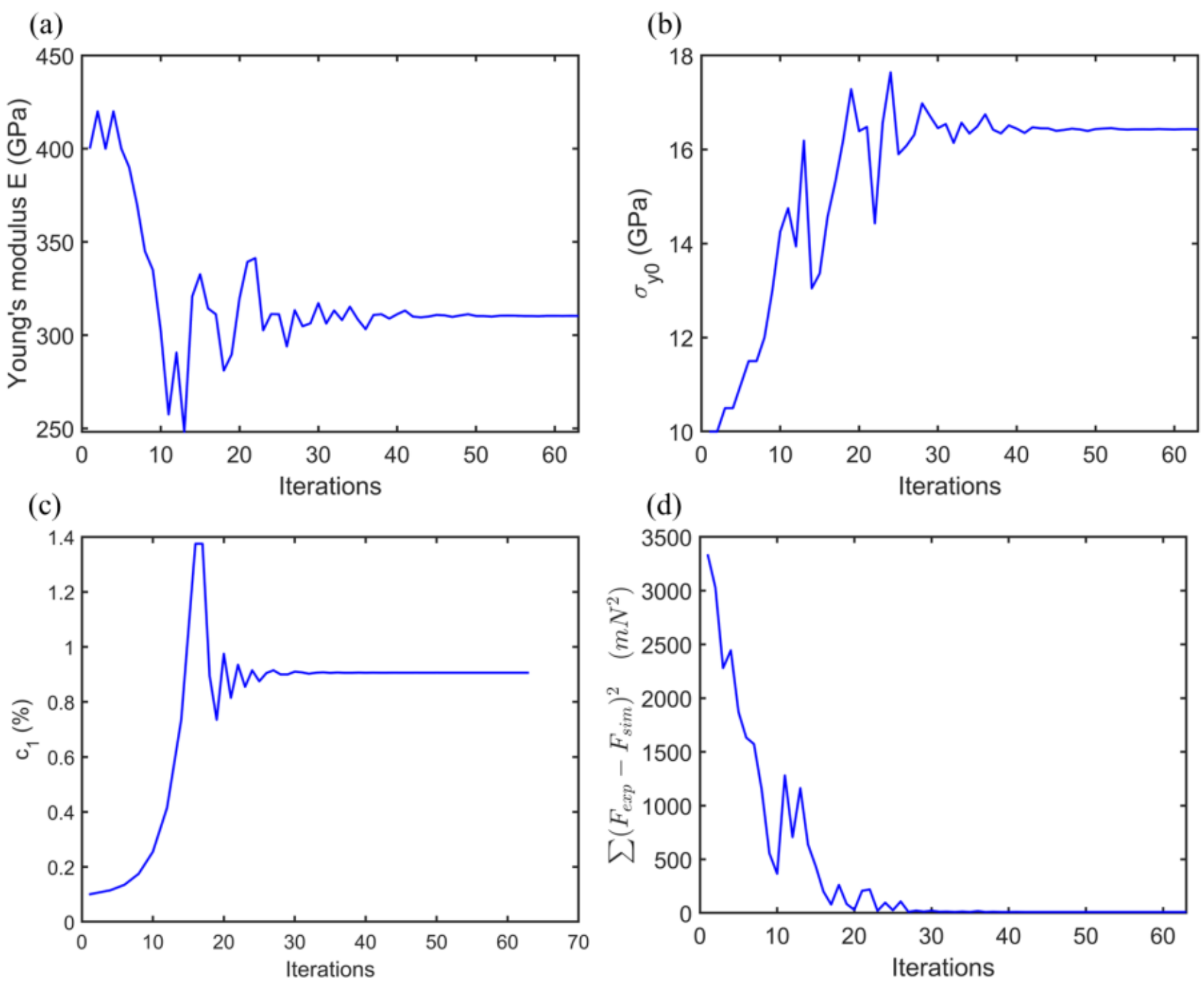

**FIG. 8.** Evolution over iterations of (a) Young's modulus $E$, (b) yield strength $\sigma_{y0}$, (c) strain amplitude $c_1$ and (d) sum of squared errors between the experimental and simulated force values.

We briefly describe the evolution of the estimated parameters throughout the optimization process. Young's modulus, initially set to 400 GPa, progressively stabilizes around 310 GPa. Similarly, the yield strength, which was arbitrarily initialized to 10 GPa, converges to a final value of 16.4 GPa. These values fall within the range of values reported in the literature from 2D FEM analyses (E = 304 GPa and $\sigma_0$ = 27 GPa) [25]. The strain magnitude parameter $c_1$, used in the constitutive model, started as 0.1% and converged to 0.91% (further insights into the evolution of strain and stress fields and their corresponding profiles over time, at the optimal parameters $c_{optim}$, $E_{optim}$, and $\sigma_{optim}$, are presented in Fig. S3 to Fig. S6). These trends demonstrate the effectiveness of the optimization routine in refining the mechanical properties and strain magnitude to match the experimental data, as documented by the drop in error shown in Fig. 8d.

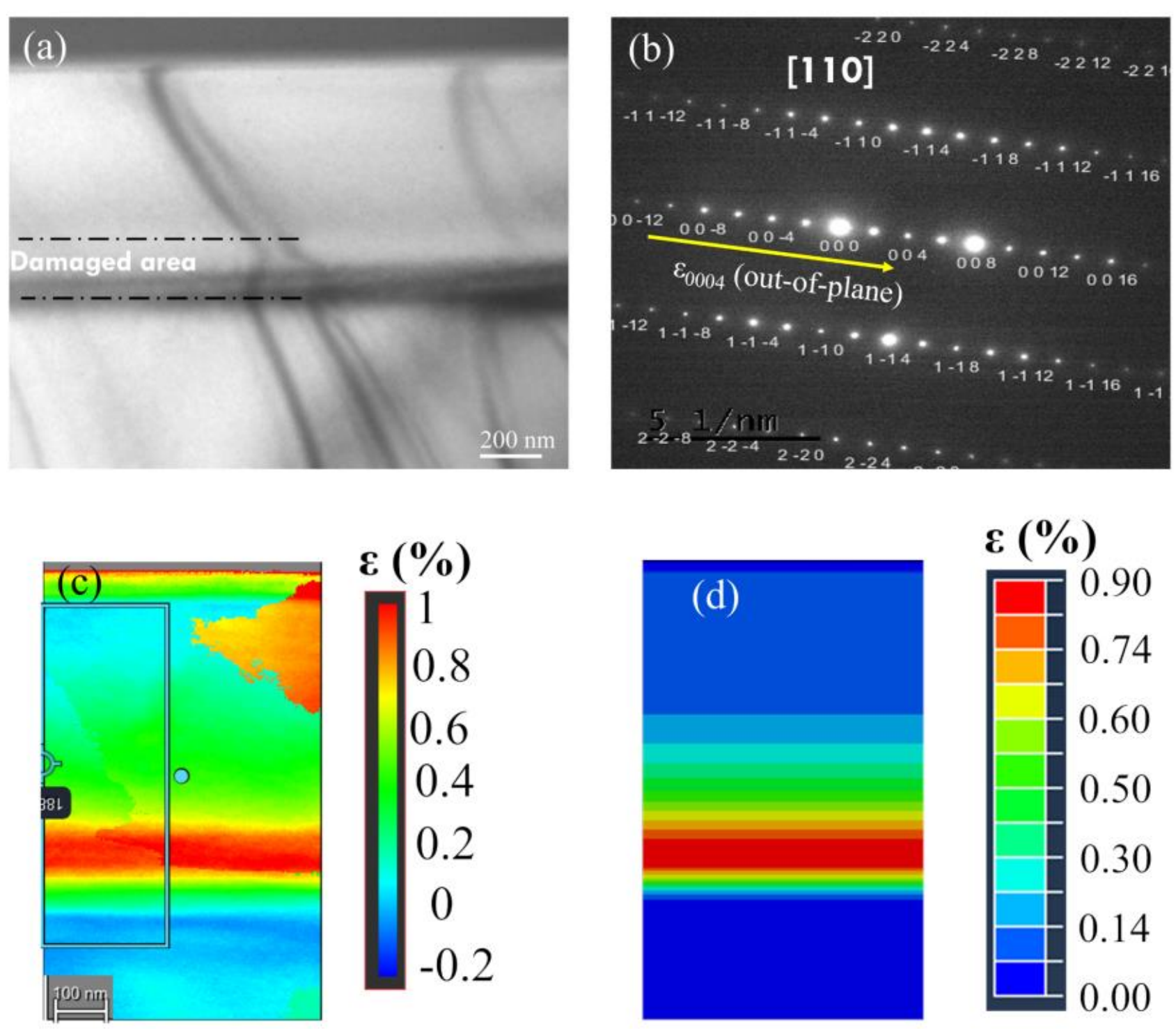

**FIG. 9.** (a) Bright-field TEM image of a cross-sectional lamella from the He- and H-co-implanted 4H-SiC sample, highlighting the damaged region near the projected ion range. (b) Selected-area electron diffraction pattern (SADP) acquired along the [110] zone axis; the (0004) reflection corresponds to the out-of-plane direction (Reproduced with permission from Daghbouj et al., Acta Materialia, 2025, 121739; licensed under a Creative Commons Attribution (CC BY) [47]). (c) Out-of-plane strain map obtained by N-PED, showing a localized tensile strain distribution with a maximum strain of ~1% near the damage peak. (d) FEM-simulated out-of-plane strain field based on the optimized strain profile and mechanical properties.

To validate the accuracy of the proposed method, experimental strain distributions were measured using N-PED in TEM and compared to the simulated strain profile. A cross-sectional TEM lamella of the He+H ion-irradiated 4H-SiC sample was prepared and imaged to identify the damage zone. The sample was aligned along the [110] zone axis, ensuring that the (0004) reflection corresponded to the out-of-plane direction, i.e., normal to the sample surface, as shown in the selected-area diffraction pattern (SADP) in Fig. 9b. For strain quantification, a reference region, far from the irradiated area and assumed to be strain-free, was selected. Strain mapping was then performed across a large field of view (Fig. 9c). The resulting strain profile shows an asymmetric Gaussian distribution, with a peak tensile strain of ~1% coinciding with the region of maximum ion-induced damage and ion concentration.

Furthermore, the strain field perpendicular to the sample surface obtained from the FEM is presented in Fig. 9d, and the corresponding strain profile is plotted along with the experimental N-PED strain data in Fig. 10a. Both the magnitude and spatial distribution of strain from the model closely align with the N-PED results, demonstrating consistency between the experimental observations and numerical predictions. The calculated sum of squared errors between the simulated and experimental force-displacement curves corresponds to a relative error of approximately 1.5%., confirming the reliability of the approach. This agreement, combined with the sensitivity analysis, supports the robustness of the extracted strain magnitude despite the inherent non-uniqueness of the inverse problem. The slight discrepancy between the estimated profile and that measured by N-PED can be attributed to strain relaxation induced by FIB cutting. This relaxation leads to a reduction in compressive stress, and consequently in strain, which is particularly evident in the profile (Fig. 10a). Moreover, the close match between the experimental and simulated force-displacement validates the robustness of the calibrated model (Fig. 10b).

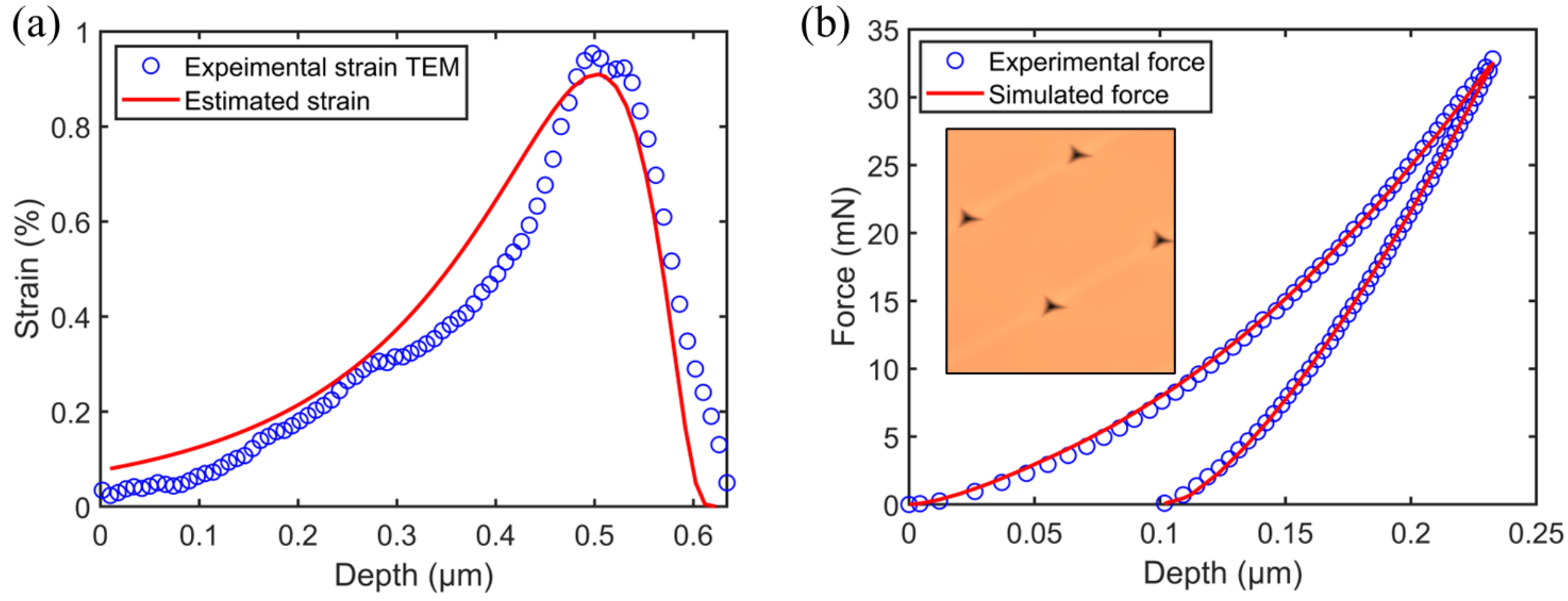

**FIG. 10.** (a) Comparison between the experimental N-PED strain profile and the FEM-implemented strain distribution along the irradiation direction, showing good agreement in both magnitude and spatial distribution. (b) Comparison between the averaged experimental and FEM-simulated nanoindentation force-displacement curves.

To investigate the interaction between irradiation-induced residual stress and indentation-induced deformation, a FIB lamella was prepared through the residual indentation impression (inset in Fig. 10b) and analyzed by TEM. The TEM image (Fig. 11a) reveals pronounced defect contrast beneath the indentation, consistent with severe plastic deformation and the presence of irradiation- and indentation-induced lattice defects, together with localized crack formation in the irradiated region. The observed cracks are attributed to the intrinsically brittle nature of SiC combined with the relatively high indentation load, which generates severe local stress concentrations beneath the indentation.

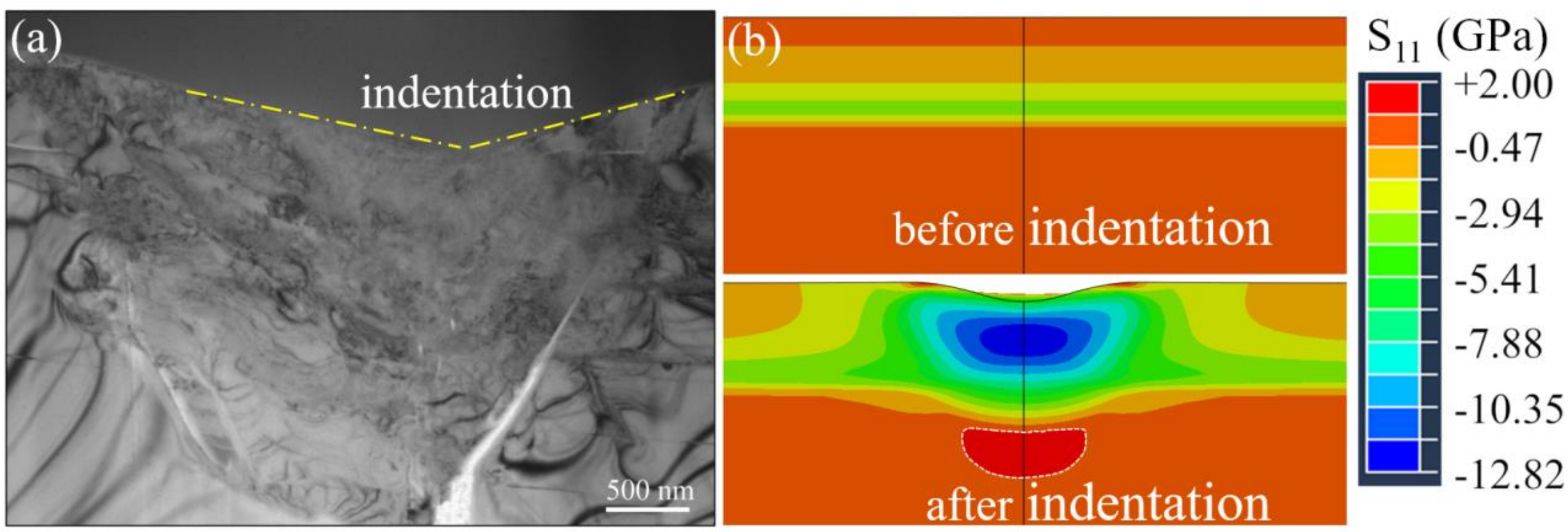

**FIG. 11.** (a) TEM image of the indented irradiated region, revealing pronounced defect contrast beneath the indentation together with localized crack formation in the irradiated region. (b) Simulated in-plane stress field (S11) before and after nanoindentation, showing the interaction between irradiation-induced residual stresses and indentation-induced stresses.

The proposed framework involves several simplifying assumptions that define its scope of applicability. First, the Berkovich indenter is approximated by an axisymmetric conical geometry, which accurately reproduces global force-displacement behavior but does not capture the local threefold asymmetry of the stress field. Second, the material response is modeled as perfectly elasto-plastic, neglecting possible strain hardening or irradiation-induced changes in fracture behavior. Third, the strain field is assumed to follow the damage profile predicted by SRIM and ERDA, which may not fully account for defect migration or stress relaxation during sample preparation. Despite these limitations, the approach remains robust for extracting effective strain magnitudes and depth distributions, as demonstrated by the close agreement with independent N-PED measurements. Consequently, the identified strain field should be interpreted as a constrained estimate of the true distribution, rather than a unique reconstruction.

To further investigate the extent and evolution of the plastic deformation zone, two-dimensional stress profiles ($S_{11}$) beneath the indenter were extracted from the FEM simulations, as shown in Fig. 11b. Before the indentation, the stress field reflects compressive in-plane stresses generated by irradiation-induced defects formed during He+H co-implantation. The resulting out-of-plane strain ($\varepsilon_{out\ of\ plane}$) is directly related to the in-plane stress through the elastic relation [60]:

$$\sigma_{\text{in plane}} = \frac{-E}{1-\nu}\varepsilon_{\text{out of plane}} \tag{8}$$

In our case, using $E$ = 310 GPa, ν = 0.17, and a peak strain of 0.91% (Fig. 10a), the corresponding compressive stress is 3.39 GPa.

The stress map before indentation shows a compressive stress peak of ~3.39 GPa near the depth of maximum irradiation damage. After nanoindentation, the interaction between the pre-existing residual stress field and the indentation-induced stress generates a more complex distribution. The compressive stress locally increases to ~12 GPa near the projected ion range, while a localized tensile region (~2 GPa) develops beneath this zone (highlighted by the dashed white circle in Fig. 11b). The strong stress gradients at the interface between these compressive and tensile regions likely promote crack initiation, consistent with the experimentally observed cracks in Fig. 11a.

Although cracking is experimentally observed, fracture is not reproduced in the FEM simulations because the present model does not include a damage or fracture criterion, such as cohesive-zone modeling or brittle crack propagation. Instead, the simulations are limited to an elastoplastic constitutive description aimed at reconstructing the irradiation-induced strain and stress fields. Incorporating fracture mechanics into the model would substantially increase the computational complexity and is beyond the scope of the present study.

The irradiation-induced tensile strain reconstructed in this study originates from the accumulation of point defects and small defect clusters generated during He+H implantation. Interstitial-type defects and dislocation loops induce local lattice dilation along the irradiation direction, while constrained lateral expansion produces in-plane compressive stresses. The resulting compressive in-plane stress increases the resistance to plastic deformation during nanoindentation, explaining the observed hardening. Importantly, this effect arises even at damage levels well below the amorphization threshold, highlighting the sensitivity of nanoindentation-based inverse methods to early-stage irradiation damage. From a nuclear materials perspective, the ability to quantify subsurface irradiation-induced strain using semi-destructive techniques such as nanoindentation provides a valuable complement to diffraction-based techniques. This approach is particularly attractive for small irradiated volumes, ion-irradiated surrogates, and post-irradiation examination, where sample availability is limited. The methodology can be extended to other nuclear ceramics and metals subjected to helium, hydrogen, or heavy-ion irradiation.

## IV. CONCLUSION

An integrated experimental-numerical framework has been developed to estimate irradiation-induced subsurface strain in single-crystal 4H-SiC subjected to sequential helium and hydrogen ion implantation. By combining depth-sensing nanoindentation with finite element modeling and a physics-constrained inverse optimization scheme, a depth-dependent eigenstrain profile is prescribed based on independently determined damage distributions, while its magnitude is identified from force-displacement data.

The optimization yields a peak irradiation-induced strain of approximately 0.91%, a Young's modulus of 310 GPa, and a yield strength of 16.4 GPa for the irradiated material. Independent N-PED measurements confirm close agreement between the reconstructed and experimentally measured strain profiles in both magnitude and spatial distribution, providing strong validation for the inverse methodology.

Beyond conventional hardness evaluation, this work demonstrates that nanoindentation, when combined with physics-based constraints, can serve as a practical and quantitative probe of irradiation-induced strain. This approach enables strain estimation in small irradiated volumes where traditional diffraction techniques are often limited, particularly in polycrystalline materials where peak overlapping occurs. Furthermore, it provides a complementary method to TEM-based strain mapping. The framework is readily extendable to other nuclear ceramics and metallic systems exposed to ion irradiation.

## SUPPLEMENTARY MATERIAL

See the supplementary material for detailed information on the fitting models and their corresponding coefficients, the implementation and workflow of the Nelder-Mead optimization algorithm, as well as the temporal evolution of the irradiation-induced strain and residual compressive stress distributions obtained from finite element simulations.

## ACKNOWLEDGEMENTS

This work was financially supported by the European Union under the project Robotics and advanced industrial production (Reg. No. CZ.02.01.01/00/22_008/0004590). CzechNanoLab project LM2023051 funded by MEYS CR is gratefully acknowledged for the financial support of the measurements/sample fabrication at LNSM Research Infrastructure. Parts of this research were carried out at IBC at the Helmholtz-Zentrum Dresden-Rossendorf e. V., a member of the Helmholtz Association.

## AUTHOR DECLARATIONS

### Conflict of interest

The authors have no conflicts to disclose.

### Author contributions

**M. Bensalem:** Investigation (equal); Methodology (equal); Software (lead); Visualization (equal); Validation (equal); Formal analysis (equal); Data curation (equal); Conceptualization (equal); Writing-original draft (equal); Writing-review & editing (equal). **N. Daghbouj:** Investigation (equal); Methodology (equal); Software (supporting); Visualization (equal); Validation (equal); Formal analysis (equal); Data curation (equal); Conceptualization (equal); Writing-original draft (equal); Writing-review & editing (equal); Supervision (lead); Resources (lead); Project administration (supporting). **J. Duchoň:** Writing-review & editing (supporting). **B.S. Li:** Investigation (supporting); Writing-review & editing (supporting); Funding acquisition (supporting). **A.T. AlMotasem:** Writing-review & editing (supporting). **S. Magalhães:** Investigation (supporting); Writing-review & editing (supporting); Formal analysis (supporting). **A. Yi:** Investigation (supporting). **F. Munnik:** Investigation (supporting); Writing-review & editing (supporting); Funding acquisition (supporting). **Xin Ou:** Investigation (supporting); Writing-review & editing (supporting). **W.J. Weber:** Writing-review & editing (supporting);

Visualization (supporting); Validation (supporting); Formal analysis (supporting). **T. Polcar:** Writing-review & editing (supporting); Visualization (supporting); Validation (supporting); Supervision (supporting); Project administration (lead); Funding acquisition (lead).

**DATA AVAILABLITY**

The data that support the findings of this study are available from the corresponding author upon reasonable request.